\documentclass[acmsmall]{acmart}
\usepackage[normalem]{ulem}
\usepackage[utf8]{inputenc}
\usepackage{mathtools}
\usepackage{alltt}
\usepackage{pifont}

\usepackage{tikz}

\AtBeginDocument{%
  \providecommand\BibTeX{{%
    \normalfont B\kern-0.5em{\scshape i\kern-0.25em b}\kern-0.8em\TeX}}}


\setcopyright{acmcopyright}
\copyrightyear{2018}
\acmYear{2018}
\acmDOI{XXXXXXX.XXXXXXX}

\usepackage{enumitem}
\setitemize{noitemsep,topsep=0pt,parsep=0pt,partopsep=0pt,leftmargin=*}

\newcommand{\bi}{\begin{itemize}}
	\newcommand{\ei}{\end{itemize}}
\newcommand{\be}{\noindent\begin{enumerate}}
	\newcommand{\ee}{\noindent\end{enumerate}}

\usepackage[labelfont=bf,textfont={bf}]{caption}

\newcommand{\OLD}{\sffamily{SNEAK}}
\newcommand{\IT}{\sffamily{niSNEAK}}
\newcommand{\SWAY}{\sffamily{SWAY}}
\newcommand{\DARK}{\cellcolor[HTML]{666666}}  
\newcommand{\MID}{\cellcolor[HTML]{CCCCCC}}
\newcommand{\LIGHT}{\cellcolor[HTML]{EFEFEF}}
\newcommand{\github}{\url{https://github.com/zxcv123456qwe/niSneak}}

\usepackage{tabularx}
\usepackage{amsmath}
\usepackage{url}
\usepackage{blindtext}
\usepackage{wrapfig}
\usepackage{graphicx}
\usepackage{algorithm}
\usepackage[noend]{algpseudocode}
\usepackage{float}
\usepackage{aliascnt}
\usepackage{color, colortbl}
\usepackage{adjustbox,booktabs,multirow}
\usepackage[skins]{tcolorbox}
\usepackage{balance}
\usepackage{xcolor}
\usepackage[skins]{tcolorbox}
\usepackage{caption}
\usepackage{lscape}

\algnewcommand\algorithmicforeach{\textbf{foreach}}
\algdef{S}[FOR]{ForEach}[1]{\algorithmicforeach\ #1\ \algorithmicdo}

\newaliascnt{eqfloat}{equation}
\newfloat{eqfloat}{h}{eqflts}
\floatname{eqfloat}{Equation}

\newenvironment{RQ}{\vspace{2mm}\begin{tcolorbox}[enhanced,width=\linewidth,size=fbox,colback=gray!15,drop shadow southeast,sharp corners]}{\end{tcolorbox}}

\begin{document}

\title{Learning from Very Little Data: On the Value of Landscape Analysis for Predicting  Software  Project Health}

\author{Andre~Lustosa}
\email{alustos@ncsu.edu}
\orcid{0000-0003-1202-3130}
\affiliation{%
  \institution{North Carolina State University}
  \city{Raleigh}
  \state{NC}
  \country{USA}
}

\author{Tim~Menzies}
\email{timm@ieee.org}
\affiliation{%
  \institution{North Carolina State University}
  \city{Raleigh}
  \state{NC}
  \country{USA}
}

\setcopyright{none}
\settopmatter{printacmref=false} % Removes citation information below abstract
\renewcommand\footnotetextcopyrightpermission[1]{} % removes footnote with conference information in first column

\definecolor{ao}{rgb}{0.2, 0.2, 0.6}

\definecolor{ao}{rgb}{0.91, 0.45, 0.32}
	\definecolor{ao}{rgb}{0.89, 0.0, 0.13}
\definecolor{aoc}{rgb}{0.0, 0.5, 0.0}

% % Note that keywords are not normally used for peer review papers.
% \begin{IEEEkeywords}
% Hyperparameter Optimization,  Project Health, Clustering, Machine Learning, Semi Supervised Learning
% \end{IEEEkeywords}}
\renewcommand{\shortauthors}{Lustosa et al.}

\begin{abstract}
When data is scarce, software analytics can make many mistakes. 
For example, consider
learning predictors
for open source project health (e.g. the number of closed pull requests in twelve months time).
The training data for this task may be very small (e.g. five years of data, collected
every month means just 60 rows of training data).
The models generated from such tiny data sets can make  many prediction errors.

Those   errors can be tamed by  a {\em landscape analysis}
that selects 
  better  learner control parameters.
Our {\IT} tool
   (a)~clusters the data to find the general landscape of the hyperparameters; then
 (b)~explores a few representatives from each part of that landscape. 
{\IT} is   both 
 faster  and more effective than 
 prior state-of-the-art   hyperparameter optimization
algorithms (e.g. FLASH, HYPEROPT, OPTUNA).

The configurations found by {\IT} have far less error than other
methods.
For example,
for project health indicators such as $C$= number of commits; $I$=number of closed issues, and $R$=number of closed pull requests,  
{\IT}'s 12 month prediction errors  
are \{I=0\%, R=33\%\,C=47\%\} while  other methods have    far larger  errors of  \{I=61\%,R=119\%\,C=149\%\}. 
We conjecture that   {\IT} works so well since it   finds the most informative regions
of the hyperparameters, then jumps to those regions. Other methods
(that do not reflect over
the landscape) can waste time exploring less informative  options.

Based on the above,  we recommend   landscape analytics (e.g. {\IT}) especially  when learning from very small data sets.
This paper  only explores the application of {\IT} to project health.
That said, 
we see nothing in principle that
prevents the application of this technique to a wider range of problems. 

To assist other researchers in repeating, improving, or even refuting our results, all our scripts and data are available on GitHub at \github.

\end{abstract}



% \keywords{Hyperparameter Optimization,  Project Health, Clustering, Machine Learning, Semi Supervised Learning}

\maketitle
% \sloppy

\section{Introduction}\label{intro}

Open source software (OSS)  projects  are most successful when they are supported by a large community
of developers.
Such support can take many important forms
including
(a)~programmer time (to fix bugs or write documentation or build APIs);
(b)~financial donations (to fund infrastructure);
or (c)~public endorsements
(e.g. when        companies join the board
of directors of OSS foundations).

 Attracting such support is  
easier when projects show healthy trends in their development. 
Many researchers comment that healthy OSS projects are characterized by much developer activity~\cite{wahyudin2007monitoring,jansen2014measuring,manikas2013reviewing,link2018assessing,wynn2007assessing,crowston2006assessing}. 
For example, Han et al. note that popular open-source software (OSS) projects tend to be more active~\cite{han2019characterization}.
Accordingly,  we seek to improve methods that   predict the future activity of an OSS project
(e.g. number of closed pull requests in twelve months time).

Sarro et al.~\cite{Sarro16} assert that for this kind of project estimation,
acceptable error rates are 30\% to 40\% (measured as {\em abs(actual - predicted)/ actual}).
But when project data is limited, 
it is hard to meet Sarro's requirements. 
For example, consider the task of learning an OSS health predictor using the 60 rows (ish) of data that can be extracted from such projects\footnote{Why 60 rows? Most of their projects have a lifespan of 3 to 7 years~\cite{xia2022predicting,shrikanth2021early}
   which meant that, for each project, they could reason about   36 to 84 rows (i.e. one row collected monthly for 3 to 7 years).}.
   As we will show below, the models learned from such small data sets have 
error rates much worse than the limits stated by   Sarro et al.

% To meet the requirements of Sarro et al.,  we 
% explore
% {\em hyperparameter optimization}.
% A common result in software analytics~\cite{agrawal2018better,agrawal2019dodge,agrawal2020better, Fu2016Tuning, fu2016differential,xia2020sequential,tu2021frugal,debtfree,9226105} (and other domains outside of SE~\cite{bergstra2011algorithms,storn1997differential,nair2018finding,chen2018sampling,akiba2019OPTUNA,9463120}) is that automatic hyperparameter optimization  algorithms can find
% good  configuration settings.
 We show here that a new   optimizer called {\IT}, based on 
{\em  landscape analytics},  can dramatically reduce prediction errors, even when learning models from very little data.
{\em Landscape analytics}~\cite{Malan21,bosman2020visualising,mersmann2011exploratory,ochoa2008study,belaidouni1999landscapes} 
maps out the space of data being explored  in order to find the most information-rich part of the data.
Once that is known, then we can leap to the more informative parts of the problem space.  Most landscape analytics methods require a pre-enumeration of most of the   space~\cite{Malan21}. But in the case of predicting for open source project health, this is impractical.
For example,  Table~\ref{tab:paramGrids}
lists   960,000 different ways a learner might be configured for predicting open source project health.   Assuming five seconds per learner (which might be an underestimate), then a study of all 960,000 options,  repeated 20 times (to check for
external validity) would terminate after
 \mbox{$960,000 * 5 /(3600*24)*20 = 1,111$ days}
(and note  that if Table~\ref{tab:paramGrids}  explored more options
or more learners, or slower methods such as deep learning, then those runtimes would get even longer.)

\begin{table}[!t]

 \small
\centering
\begin{tabular}{r|rrl|r}
\textbf{Parameter}               & \textbf{Min}         & \textbf{Max}         & \textbf{Step}       & \#options \\ \hline
\textbf{n\_estimators}           & 10                   & 200                  & 10                  & 20 \\
\textbf{min\_sample\_leaves}     & 1                    & 20                   & 1                   & 20\\
\textbf{min\_impurity\_decrease} & 0                    & 10                   & 0.25                & 40\\
\textbf{max\_depth}              & 1                    & 20                   & 1                   & 20\\ 
\textbf{criterion}               & \multicolumn{3}{c|} {one of: [squared, absolute, poisson]}         & 3\\\cline{2-5}
\multicolumn{3}{c}{~} & Total (20*20*40*20*3): & 960,000
\end{tabular}
\caption{Options for ensembles of regression trees.}
\label{tab:paramGrids}
\end{table}
Landscape
analytics
  can reduce the cost of that search. For example,
{\IT} can find   configurations that result in low errors after looking at around 100 options
(while other tools needed to examine thousands
more options~\cite{akiba2019OPTUNA,bergstra2011algorithms}).
Not only that, but
{\IT}'s
predictions have errors rates within the Sarro et al. limits. In fact, when
predicting for one particular project health indicator (number of closed
issues), we found a median error rate of 0\%\footnote{But note that  30\% of the time, error rates up to 30\% were observed.}.
We conjecture that those
other optimizers performed worse
since they used a somewhat uninformed
search based on  random mutations.
On the other hand, 
{\IT} works well
 because  it carefully reflects on  the shape
of the data before deciding where to go next.

This paper explores three research questions:
\bi 
\item 
\textbf{RQ1 -}  {\em Does small sample size damage our ability to perform project health prediction?}. This is our {\em baseline} question. 
The central motivation of this paper is that   small
sample sizes in analytics can lead to errors in project health estimation. Hence, before this paper does anything else, we need to back up that claim. What we will show as part of {\bf RQ1} is that many methods in common use perform very badly on our tiny data sets.
\item
 \textbf{RQ2 -} {\em Can landscape analysis mitigate for small sample sizes? }
Here, we compare the better results from {\bf RQ1} (that make no use of landscape analysis)
with the new methods of this paper (that use landscape analysis).  As shown in the {\bf RQ2}
results, adding landscape analysis significantly improves our ability to make predictions about
very small data sets. 
\item

 \textbf{RQ3 -} {\em How much landscape analysis is enough?} 
 This question is somewhat technical and  addresses some internal design
 decisions. {\IT} runs in multiple {\em sweeps} across the landscape where each sweep  focuses on a different part of the data
 (e.g. one sweep only looks at the dependent variables
 while another only looks at the independent values). 
This raises the question of whether we are ``sweeping''  too much or too little. As seen by the results
of {\bf RQ3},      some sweeping is better than no sweeping, and too much is contraindicated.
 \ei
The rest of this paper is structured via
  guidelines from Wohlin \& Runeson et al.~\cite{wohlin2012experimentation} on empirical software engineering. We present our methods in \S\ref{algo}, define and explain our case study and in \S\ref{methods}, then show results in \S\ref{results}.  See also \S\ref{threats} for a discussion on the threats to the validity of our conclusions.

\section{Background}\label{background}

\subsection{Motivation} 
\begin{figure}[!t]
\begin{center}
\includegraphics[width=.6\textwidth]{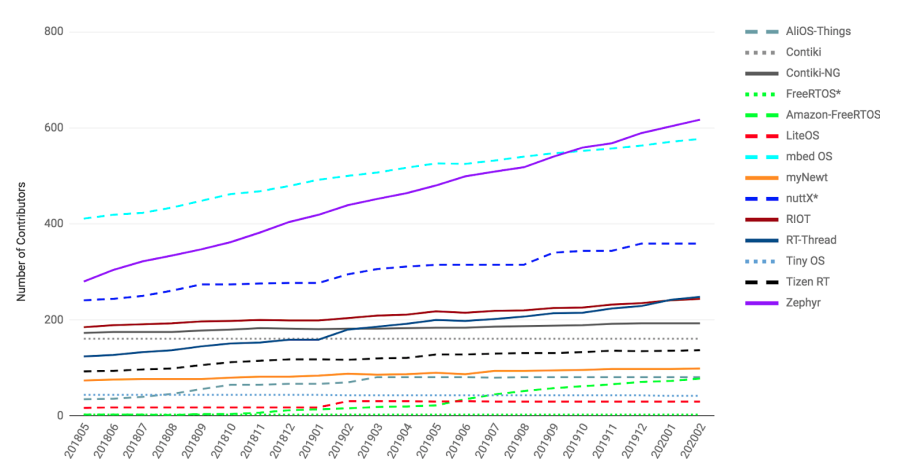}
\end{center}
\caption{Trends in number of contributors across 14 real-time OS projects since
2018. Zephyr (purple) has out-paced other similar software projects.}\label{zep}
\end{figure}
\begin{table}[!b]
\scriptsize
\begin{center}
\begin{tabular}{r | l}
Feature &Description\\\hline
dates &The end date of monthly data collection\\
monthly commits& Total number of commits created in last month\\
monthly commit comments &Total number of commit comments created in last month\\
monthly contributors &Total number of contributors that at least have one commit in last month\\
monthly open PRs &Total number of pull requests opened in last month\\
monthly closed PRs &Total number of pull requests closed in last month\\
monthly merged PRs& Total number of pull requests merged in last month\\
monthly PR mergers &Total number of pull requests mergers in last month\\
monthly PR comments& Total number of pull request comments created in last month\\
monthly open issues &Total number of issues opened in last month\\
monthly closed issues &Total number of issues closed in last month\\
monthly issue comments &Total number of issue comments created in last month\\
monthly stargazer& Total number of new stars acquired in last month\\
monthly forks &Total number of forks occurred in last month\\
monthly watchers& Total number of new watchers acquired in last month
\end{tabular}
\end{center}
\caption{Project health indicators. From Xia et al. ~\cite{xia2020predicting}.}\label{healthind}
\end{table}

 This section explains our motivation for exploring   open source project health. 

In summary, in many scenarios, it is important to predict future health (or, if that health is declining,
to take steps to address that issue). 
For example,
many commercial companies use open-source packages  in the products they sell to customers.  
Those companies seek projects that they   trust will remain viable for the foreseeable future.

Also, projects with an unhealthy reputation lose their  community support. This is not ideal
since
OSS projects grow when they attract developers.
Large OSS projects (e.g. the LINUX or APACHE foundation) have board members representing organizations that allocate considerable resources to maintaining and extending   software. 
 Attracting such investments is far easier when  projects can show potential investors   healthy trends in their project development.
For example, in the Apache ecosystem, there are 14 real-time time operating systems. 
Measured in terms of code commits per month, Figure~\ref{zep} shows that most of these are stagnant (e.g. LiteOS and myNewt),
while the remaining two (Zephyr and mbed~OS) show a strong increase in the number of commits per month.

 There are many ways to document those trends
and, for our research,
we use a list of trend indicators
selected by a survey by    hundreds of decision-makers from
dozens of open source projects~\cite{xia2022predicting}.
In that survey, it was    found that
 decision-makers  wanted methods to predict
{\em open source health indicators} of Table~\ref{healthind}
such as  {\em commits};  {\em closed pull requests}; and  {\em number of contributors}.

 \begin{wrapfigure}{r}{3in}
\caption{Importance of indicators to   OSS developers. PR=pull requests.
From~\cite{xia2020predicting}. }
\label{tbl:survey}
\begin{adjustbox}{max width=0.98\textwidth}
\includegraphics[width=3in]{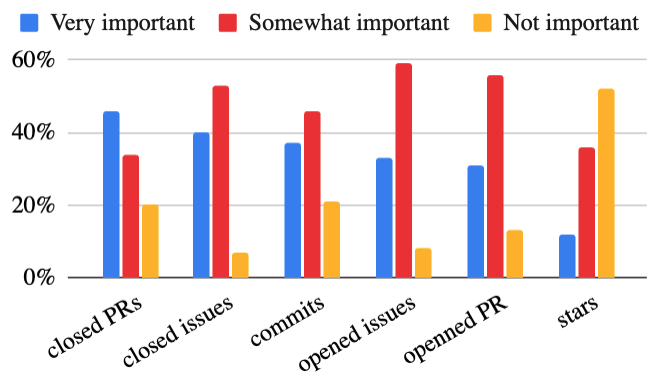}
% \footnotesize
% \begin{tabular}{lc|c|c}\cline{2-4}
% & \multicolumn{3}{|c|}{Important?}\\
% \rowcolor{gray}
%  &\textcolor{white}{ Very}   & \textcolor{white}{ Somewhat }  & \textcolor{white}{ Not }  \\ \hline
 
%   closed pull-requests   & 46\% & 34\% & 20\% \\

%    closed issues         & 40\% & 53\% & 7\% \\

%    contributors          & 33\% & 46\% & 21\% \\

%   opened issues         & 33\% & 59\% & 8\% \\

%    opened pull-requests & 31\% & 56\% & 13\% \\
%   stars.                & 12\% & 36\% & 52\%
% \end{tabular}
\end{adjustbox}
\end{wrapfigure}  Figure~\ref{tbl:survey} shows the results when Xia et al.~\cite{xia2020predicting}  asked  developers  to rank the indicators.
That survey sought the opinion of ``core developers''; i.e. those who make pivotal decisions about software.  
Xia et al. found 112 core contributors from 68 projects and asked them ``what product health indicators would prompt them to take action within their project?''.  
Some indicators were not so important (e.g. number of stars assigned to a project in GitHub) while others were seen to be far more valuable (e.g. closed pull requests). 
 
The rest of this paper explores methods for predicting the items  in Figure~\ref{tbl:survey} that were ranked largest in terms
of ``very important''  
 (a)~number of closed issues, and (b)~number of closed pull requests; and (c)~number of commits.

 \subsection{Differences to Related Work}\label{related}

Before   going further, we digress here to     distinguish this work from prior research.

Traditionally,   in the age of waterfall projects,
 project effort predictions tried to guess the funds  required for {\em project completion}.
 To do so, some nominal estimate was generated (which was then inflated according to known risk factors inside a project). 
For example, Boehm et al.'s COCOMO model~\cite{chulani1999calibrating} (developed in 1981 and updated in 2000)  generated estimates of total project effort (measured in terms of 150-hour work-months) that were within 30\% of actual, 71\% of the time.  
Robbles et al.~\cite{robles2022development} report that for a large class of OSS projects,  developers will spend much of their time working full time on a single project. 
Hence, for those projects, it is possible to estimate the total effort that will be spent on a project, early in its development using (as shown by Robbles et al.)
project features  extracted from an online version control systems (e.g. GitHub).

Robbles'  work inspired Xia et al~\cite{xia2020predicting} to try  applying the Boehm  methods to OSS teams; specifically the APACHE and LINUX kernel development teams.
Xia et al. found that those teams expressed little interest in estimates of ``effort to finish'', since no one in these teams expected to ever ``finish'' their projects (those developers are funded in perpetuity to maintain and extend their tools).  

Since Xia et al. had so much trouble with waterfall-style estimation, in this work we explore health indicators that can be applied to on-going projects.
Many researchers have explored such health indicatiors~\cite{bao2019large,jarczyk2018surgical,kikas2016using,han2019characterization,qi2017software,chen2014predicting,wahyudin2007monitoring,jansen2014measuring,manikas2013reviewing,link2018assessing,wynn2007assessing,crowston2006assessing}.   For example,  Bao et al.~\cite{bao2019large} apply  Naive Bayes, SVR, Decision Tree, KNN and Random Forest  to 917 projects from GHTorrent to predict long term contributors (which they determine as the time interval between their first and last commit in the project is larger than a threshold.). They found  random forest achieves the best performance.
Kikas et al.~\cite{kikas2016using} built random forest models to predict the issue close time on GitHub projects, with multiple static, dynamic and contextual features. They report that the dynamic and contextual features are critical in such prediction tasks.
Jarczyk et al.~\cite{jarczyk2018surgical} used generalized linear models for predictions of issue closure rate. Based on multiple features (stars, commits, issues closed by the team, etc.), they found that larger teams with more project members have lower issue closure rates than smaller teams, while increased work centralization improved issue closure rates. 
Other developing related feature predictions also include the information on commits, which is used by Qi et al.~\cite{qi2017software} in their software effort estimation research of OSS projects, where they treat the number of commits as an indicator of human effort.
Chen et al.~\cite{chen2014predicting} used linear regression models on 1,000 GitHub projects to predict the number of forks; they concluded that this prediction could help GitHub recommend popular projects and guide developers to find projects that are likely to succeed and worthy of their contribution.

There are several important differences between our work and the above.
For the most part, the above explored a single measure of project health (and not the many measures explored here). Also, their data collection methods were somewhat opportunistic in that they seem to have found a data source, then taken steps to exploit that data. We say this since when
we went looking for their kinds of data, we often found that the available information
was somewhat different to that used in the above. Like Xia et al., we found that, for many projects,  we could reliably collect 
information about the Table~\ref{healthind} items. Hence, we base our predictors on those attributes. 

Another difference of our work to the above is that,
with the exception of Bao et al.~\cite{bao2019large},
most of the prior work used data miners    {\em without} hyperparameter optimization. The importance of
hyperparameter optimization is discussed in the next section.

\subsection{Hyperparameter Optimization}\label{hpola}

When a learner executes, it uses many decisions about what things to explore and what things to ignore. 
Those decisions are controlled by the {\em hyperparameters} of the learner.
Example hyperparameters include (a)~how many neighbors to use in nearest neighbor classification or (b)~how many trees to include in a random forest or (c)~how to configure the architecture of a deep learner.

\newcommand{\GG}{\cellcolor{gray!15}}

Hyperparameter optimization is a very active field of research. A common result in software analytics~\cite{agrawal2018better,agrawal2019dodge,agrawal2020better, Fu2016Tuning, fu2016differential,xia2020sequential,tu2021frugal,debtfree,9226105} (and other domains outside of SE~\cite{bergstra2011algorithms,storn1997differential,nair2018finding,chen2018sampling,akiba2019OPTUNA,9463120}) is that automatic hyperparameter optimization   can find
settings that dramatically improve some algorithm.
We list below some of the algorithms commonly seen in the literature.  
Our reading of the literature is that, of the following, HYPEROPT and OPTUNA represent the prior state-of-the-art. 
One complaint we have against the Xia et al. analysis is that they made scant use of those state-of-the-art tools (their paper makes only one small study with HYPEROPT, on limited data).

%  \begin{table}[!t]
% \caption{Median MRE performance deltas between off-the-shelf and tuned predictors for project health indicators, one month into the future. From~\cite{xia2020sequential}.
% All these values are {\em statistically significantly different}, and {\em worse}, than the performance achieved by Xia et al.'s hyperparameter optimization methods.
% Cells in gray show deltas larger than the Sarro-limit of 40\%~\cite{Sarro16}.
% KNN=Nearest Neighbors, SVR=Support Vector Regression, LNR=Linear Regression, RFT=Random Forest and CART=Regression Trees.
% }\label{limit}

% \small
% \begin{tabular}{r|rrrrr}
% prediction,  &  KNN &LNR &SVR &RFT &CART  \\ \hline
% commit      &  +15\%& \GG +112\% &\GG +65\% & \GG +42\% & +18\% \\
% closePR      & +23\%&\GG +44\% &+14\% & +24\% & +14\% \\
% closedISSUE  & +35\%& \GG +45\% &\GG +67\%& +21\%& +25\% 
% \end{tabular}
% \end{table} 

\subsubsection{Grid Search (GS)~\cite{bergstra2011algorithms}}
As described by Bergstra et al.~\cite{bergstra2011algorithms}, GS defines all hyperparameter options in a set of vectors. 
GS then executes over the cross product of those options. Note that GS will execute all combinations, and as such for limiting CPU usage one must define a smaller search.  

\subsubsection{Random Search (RS)~\cite{bergstra2011algorithms}}

Random search starts with the same option vectors as grid search.  
But instead of evaluating all of the models in the grid, RS evaluates a random subset of these models selecting the best configuration from this subset. 
Both GS and RS have been previously used towards the optimization of models for defect prediction~\cite{nevendra2022empirical}.

\subsubsection{Differential Evolution (DE)~\cite{storn1997differential}}
Differential Evolution is an evolutionary algorithm that executes in several {\em generations}
of {\em mutate-crossover-select}. 
DE starts like random search and creates a small initial population of randomly selected configuration options (say, 10 vectors of options per feature being configured). 
For each generation, each option is  compared to a newly created option that is a mixture of three other options in the population. 
The new option replaces the old, if it is better.
After generation\#1, the invariant of DE is that each option in  a  population in generation $g$ is superior to  at least $g-1$ other options. 
Hence, as the generations mature, DE mixes together new items  built from  items of increasing  superior options.
DE was previously used for defect prediction by (e.g.) Fu et al.~\cite{Fu2016Tuning} on data from open-source JAVA systems.

\subsubsection{FLASH~\cite{nair2018finding}}
FLASH is a sequential model-based optimization (SMBO) method, introduced by Nair et al.~\cite{nair2018finding}, that outperformed the then-known state-of-the-art.  SMBO methods use a surrogate model built from options-evaluated-to-date to make guesses about all the as-yet unevaluated options.
Then (a)~the ``most interesting'' option is then evaluated, and (b)~the surrogate is then updated so that we can make better guesses in the future.  
When used in practice, FLASH's surrogate model might make 100,000s of very fast guesses in order to select the best $10^1$ to $10^2$ configurations that actually deserve evaluation.
Nair et al. used FLASH to find the best possible configurations for 7 different software systems.

\subsubsection{{\SWAY}~\cite{chen2018sampling}}\label{sway0}
The {\SWAY} oversampling approach recursively
clusters 10,000 randomly generated configurations (perhaps generated via theorem prover from domain constraints).
{\SWAY} is a greedy-search algorithm that always commits to the topmost partition  (without checking if any other partition is better).
Later in this paper, we will discuss (\S\ref{better}) the Zitzler domination predicate
and other details present in {\SWAY}.
Previously, in software engineering,
{\SWAY}~\cite{chen2018sampling} was applied towards the optimization of software process models and software product lines.

\subsubsection{HYPEROPT~\cite{bergstra2015hyperopt}}
HYPEROPT is a hyperparameter optimization framework introduced in 2015 as a Python package.
At the time of this writing, the two papers that lead to HYPEROPT~\cite{bergstra2012random,bergstra2011algorithms} have been cited 7668 and 3543 times, making this the most mentioned algorithm in this survey. 
At its core, it uses either RS or the Tree-Parzen Estimator (TPE) as the optimization algorithm. TPE is an SMBO algorithm introduced by Bergstra et al.~\cite{bergstra2011algorithms}, that will model the sequential model optimization via the use of non-parametric statistical densities. 
Yedida et al.~\cite{agrawal2021simpler, yedida2021value} have used HYPEROPT for software defect prediction.   
Classic HYPEROPT does not allow for multi-objective optimization so when we use it, we will just tune to minimize the magnitude of relative error (MRE) values.

\subsubsection{OPTUNA~\cite{akiba2019OPTUNA}}
OPTUNA is a more recent hyperparameter optimization framework than HYPEROPT that uses
MOTPE~\cite{ozaki2020multiobjective}, a more advanced version of TPE. 
MOTPE allows for the optimization of multiple performance metrics at once by analyzing the results of the current models against the currently observed frontier of solutions via using a function that calculates the expected hypervolume improvement at each step of its optimization.
To the best of our knowledge, OPTUNA has not been used extensively for
software engineering problems (but it has been used in domains such as disease diagnosis~\cite{zhou2022predicting}, equipment performance forecast~\cite{mashlakov2019hyper} and others). 
We assert that it is important to baseline our new methods against OPTUNA (and MOTPE) since this is a clear extension and improvement of HYPEROPT.

\subsection{Landscape Analytics}\label{latics}

The experiments of this paper show that, at least for predicting future trends in OSS projects, we can outperform {\em all} of the above hyperparameter optimization
methods through {\em landscape analytics}.  
This section describes that kind of analytics.

To understand landscape analytics, consider a function:
\begin{equation}\label{one}Y=f(X)\end{equation}
where $X$ contains one or more {\em independent features} and $Y$ contains one or more {\em dependent features}. 
If we cluster together with the $X$ points or the $Y$ points, we can draw out the shape of the data. 
We call this shape the ``landscape''.  

The process of data mining can be characterized as a search across landscapes.
Given many examples of  $(X_1,Y_1),(X_2,Y_2),..\mathit{etc}$ then a  learner $L$ seeks some model $f$ that knows where parts of the $X$ landscape connect to particular parts of the $Y$ landscape.
Also, a learner may contain hyperparameters $W$ about (e.g.) how many neighbors to use in the nearest neighbor classifier or (e.g.) how many times to divide numerics. 
Different hyperparameters will result in different models, so we must rewrite Equation~\ref{one} as:
\begin{equation}
Y=L(W)(X)
\end{equation}
To further guide that search of the hyperparameters, we note that in most domains, there is some preference function $P$ that tells us which parts of the dependent landscape are better than others. 
E.g. if $Y$ contains some measure of false alarms, then {\em lower} false alarms might be scored {\em higher}.  
Within this framework, the goal of hyperparameter optimization is to find the hyperparameters $W$ that lead to dependent variable settings that maximize the score:
\newcommand{\argmax}{\mathop{\mathrm{argmax}}}  
\begin{equation}
\argmax_{w_i  \subseteq W} \left(P(L(w_i)(X))\right)
\end{equation}
The list of all $W$ configurations also forms a landscape (separate to the $X,Y$ landscapes).
{\em Landscape Analytics} is the process of understanding and then exploiting the shape of these landscapes. Not all landscapes are ``cliffs'' where
the data changes suddenly and steeply. 
When changing many options has similar effects,
the landscape can be quite smooth and easily
mapped out with just a few samples. 
Hence, many methods~\cite{liu2016comparative,van2020survey,ros2017machine,mit06,h2022}  reason by cluster the data,  then explore just a few representatives from each cluster. 
To say that another way, when the landscape is smooth,   it is not necessary to search everywhere. 
Rather, it can be (much) faster to:
\be
\item  Look  around that space with just a few samples;
\item Then leap over to regions of denser information.
\ee
While it is not widely acknowledged, in our opinion, landscape analytics generalizes a range of algorithms from different fields:

\begin{table}[b!]
\footnotesize
\begin{tabular}{l|l|c|l} 
\rowcolor{black!50}  \textcolor{white}{\textbf{Assumption}} &\textcolor{white}{Notes} & \textcolor{white}{\# of papers} & \textcolor{white}{Example} \\ \hline
    \textcolor{ao}{\bf Binary space  }
    & All features have arity=2 and are often used in a theorem prover.                          & 11  & HDBL~\cite{belaidouni1999landscapes}\\
\rowcolor{gray!15}
    \textcolor{ao}{\bf Fully enumerable space }  & The entire search space can be pre-enumerated and cached.                          & 12  & LON~\cite{ochoa2008study}\\
 \textcolor{ao}{\bf Continuous space  }                        &  Useful  for (e.g.) gradient descent. Discrete features not allowed.              & 5  & ELA~\cite{mersmann2011exploratory}\\
\rowcolor{gray!15}
 \textcolor{ao}{\bf Knowledge of optima  }   
    & The value of the achievable best solution already known.            & 10  & Static-$\phi$~\cite{whitley1995hyperplane} \\
      \textcolor{ao}{\bf Evalaution is fast/cheap  }              & Optimizers can call the evaluation function 1,000,000s of times.  & 11  & LGC~\cite{bosman2020visualising} 
\end{tabular}
\caption{Assumptions made in landscape analytics methods  surveyed by Malan~\cite{Malan21}.}
\label{tab:landscapesummary}
\end{table}

\be
    \item When used on the $X$ shape, landscape analytics  might be called ``clustering''.
    \item When used on the  $X,Y$ shape,     limited sampling might be called ``semi-supervised learning'';
    \item Similarly, in a joint analysis of the $X,Y$ shape, if we bias our ``leaps'' towards regions that (in the past) had good $Y$ scores, landscape analytics  might be called ``reinforcement learning''.
    \item And if we use landscape analytics to jointly explore the $W,Y$ shapes, then this could be called ``hyperparameter optimization''.
\ee
In a recent survey of landscape analytics methods, Malan~\cite{Malan21} notes
that the term ``landscape''
is a generalization of Sewell's 1932
concept of ``fitness landscape''~\cite{wr32}  which was first defined for
evolutionary biology.
Malan reports no less than 33 ``families'' of landscape analytics methods. Most of these methods
tend to enumerate the  whole landscape, before exploring it.
This approach is seen in many parts of the Table~\ref{tab:landscapesummary} summary of
the Malan survey; e.g.:
\be
    \item Methods that assume that    
    \textcolor{ao}{\bf evaluation is fast or cheap} usually exploit that property to make many samples across the landscape.
    \item Methods that assume
     \textcolor{ao}{\bf knowledge of optima}
      also assume that the search space has been pre-explored (to find that best point).
    \item Methods that require a
       \textcolor{ao}{\bf fully enumerable space}, by definition, must generate the whole space before reasoning can begin.
\ee
For hyperparameter optimization, it is problematic to explore and cache a large part of the search space. 
This is especially true when conducting large experiments and/or exploring slower data mining algorithms (e.g. deep learners).
Quite apart from CPU issues, there are other reasons we need to look beyond standard landscape analytics tools. 
As shown in Table~\ref{tab:landscapesummary}:
\bi
    \item Many current landscape analytics methods require
    \textcolor{ao}{\bf   knowledge of optima}.
    For many data mining applications, the achievable maximum is not recall=1, false alarm=0 but some point struggling to achieve (but not reaching) that zenith.
    \item Many current landscape analytics methods cannot handle features of mixed types since they rely on (e.g.) gradient descent over a 
         continuous space
       or (e.g.) theorem proving executed on a \textcolor{ao}{\bf binary space}.   
    This is problematic for hyperparameter optimization since the policies that control data miners are often a combination of numeric and discrete choices.
\ei
For these reasons, we built our own landscape analytics tool, differing from existing ones   with the following properties:
\bi
    \item It can explore heterogeneous data (discrete and symbolic data) using simple discretization algorithms.
    \item It only enumerates a subset of the $W,X$ hyperparameters and independent features.
    \item It only enumerates a small subset of the $Y$ independent features.
\ei
That tool is called {\IT} and is described in the next section.

\section{About {\IT} }\label{algo}

{\IT} is based on Chen et al.'s {\SWAY} algorithm~\cite{chen2018sampling}. It takes a binary dataset comprising of multiple configurations as an input and outputs (a) a binary tree of clustered configurations, from which it reports (b) the best configuration.
% XXX strange stand out
% {\SWAY}, originally, compared two approaches to optimization:
% \begin{itemize}
% \item
% A {\em traditional approach} that mutated (say) $10^2$  individuals across (e.g.) $10^2$  generations;
% \item
% The SWAY {\em oversampling approach} that  builds generation$_0$with  10,000, individuals,
% which are then recursively clustered and pruned (by removing sub-trees whose roots are   dominated by a near neighbor).  
% \end{itemize}
While an interesting prototype, {\SWAY} suffers from some serious drawbacks.
Both SWAY and {\IT} recursively
partition the data, but {\SWAY}'s greedy search always commits
to the top-most partition  (without checking if any other partition is better), whereas
{\IT}  takes a global approach.
Specifically, after all the data is clustered, {\IT} reflects over all the nodes to find which sub-clusters ``best'' split the data
(and ``best'' means ``splits the data such that some attribute range most separates the two splits'').

As such we include {\SWAY} as a baseline algorithm when performing this hyperparameter optimization task.

\subsection{Preliminaries}
To understand {\IT},  , we must first
what it means to  
``evaluate'' in order to find ``better'' options.

\subsubsection{``Evaluation''}

{\em Evaluation}
means configuring a learner with a set of hyperparameters, running it on some training data,
then applying the resulting model to test data.

\subsubsection{``Better''}\label{better}
The result of such an {\em evaluation} is a set of  performance measures associated with a working learner. As shown in   Table~\ref{evals}, we have four such measures (Pred40, SA, MRE, D2H). When dealing with multiple goals,
a 
{\em domination} predicate is required
        to decide if one set  of performance measures
        $A=\{\mathit{Pred40}_i, \mathit{SA}_i, \mathit{MRE}_i, \mathit{D2H}_i\}$
        is ``better''   than $B=\{\mathit{Pred40}_j, \mathit{SA}_j, \mathit{MRE}_j, \mathit{D2H}_j\}$. 
Two such predicates are {\em boolean domination} and the {\em Zitzler multi-objective indicator}~\cite{zit02}.   Boolean domination says one thing is better than another if it is no worse on any single goal and better on at least one goal. 
We prefer Zitzler to boolean domination since we have a  four-goal optimization problem, and it is known that boolean domination often fails for two or more goals~\cite{Wagner:2007, Sayyad:2013}. 
Zitzler favors
$B$ over $A$  model if jumping to $A$ ``loses'' most:
\bi
\item Let$ \textit{worse}(A,B)  = \textit{loss}(A,B) > \textit{loss}(B,A)$
\item Let
$ \textit{loss}(A,B) = \sum_{j=1}^n -e^{\Delta(j,A,B,n)}/n$
\item Let 
       $ \Delta(j,A,B,n)  =  w_j(o_{j,A}  - o_{j,B})/n$
       \ei
where ``$n$'' is the number of objectives (for us, $n=4$) and $w_j\in \{-1,1\}$ depending on whether we seek to maximize goal $x_j$ and $o_{j,A}, o_{j,B}$ are the scores seen for objective $o_j$ for $A,B$, respectively.

\begin{table}[!t]\small
\begin{tabular}{|p{.95\linewidth}|}\hline
We assess our project health predictors via widely-used criteria from the literature~\cite{reddy2010software, Sarro16, shepperd2012evaluating}.

\bi
\item
{\em MRE} is the ``magnitude of relative error'' and can be computed from  {\em abs(actual - predicted)/ actual}.
\item
Pred40 assesses the Sarro limit    and is defined as 
   $ \mathit{Pred40} =  \mathit{Count(MREs,} \leq 40)/||\mathit{MREs}||$
 \item
Standardized accuracy (SA) baselines the observed error against some reasonable, fast, but unsophisticated measurement. 
    SA is based on the Mean Absolute Error (MAE)
    i.e. $\mathit{MAE} = \frac{1}{N}\sum_{i=1}^{n} |\mathit{PREDICT}_i - \mathit{ACTUAL}_i|$ where $N$ is the size of the test set used for evaluating performance. 
    SA then is defined as the   ratio 
 $\mathit{SA} = (1 - (\mathit{MAE}/\mathit{MAE}_\mathit{guess})) \times 100$
    
    \item
    $\mathit{MAE_{guess}}$  (used in  {\em SA}) is the $\mathit{MAE}$ of a set of guesses. 
   This paper follows the standards defined by Xia et al.~\cite{xia2018hyperparameter}; i.e.   $\mathit{MAE_{guess}}$  is the median of previous months' guesses.
\ei
We seek to
  {\em maximize} Pred40 and SA and 
  {\em minimize} MRE and D2H.
  
  D2H 
(distance to heaven)  takes all the evaluated examples and asks how close each example is to the best in that set.   Given a vector $Z$ containing all   options ranked from best to worst, the D2H of a   configuration $s$   at   index $i$ is
$ \mathit{D2H}_{s} = i / |Z|$
(and the he closer we get to ``heaven'', the smaller D2H and   more we are succeeding on all       goals). To rank the vectors,  we use the Zitzler indicator of  \S\ref{better}.\\\hline
 \end{tabular}
 \caption{``Better'' = \{MRE, Pred40, SA, D2H\}.}\label{evals}
\end{table}

Some of the performance measures of Table~\ref{evals} require extra evaluations. 
As discussed below, {\SWAY} only needs to evaluate $E=2\sqrt{N}$ options to identify    good options. But  to {\em test} if {\SWAY} found options better than (e.g.) {\IT},
we     have to look for any   good options missed by {\SWAY}. Hence, for certification purposes, we     have to evaluate   other 
options, looking for anything overlooked by {\SWAY}. Accordingly, we   distinguish two evaluation counters:  $E$ and $E^+$. Here,  $E$ is the number of evaluations made by the inference procedure and $E^+$ are all the extra evaluations used to test if one algorithm is better than another (e.g. when applying the D2H measure of  Table~\ref{evals}). 
The important point about $E,E^+$ is that  suppose researchers (like us) use $E^+=N$  of some data   $D_i$ to certify that {\IT} is a useful approach. After that,
when practitioners use this algorithm on their own data  $D_{j\not=i}$, they will only need to make $E \ll E+$ evaluations.

Having documented  these counters,  
 for the rest of this paper, when we say ``number of evaluations'', we are  reporting $E$  (and not $E+$).

\newcommand{\FUNCT}{\color{blue}{FUNCTION}}
\newcommand{\INP}{\color{blue}{INPUT}}
\newcommand{\OUT}{\color{blue}{OUTPUT}}
\newcommand{\CODE}[1]{\bfseries{\sffamily{#1}}}
\newcommand{\IF}{{\CODE if}}
\newcommand{\THEN}{{\CODE then}}
\newcommand{\ELSE}{{\CODE else}}
\newcommand{\AND}{{\CODE and}}
\newcommand{\FOR}{{\CODE for}}
\newcommand{\IN}{{\CODE in}}
\newcommand{\DO}{{\CODE do}}
\newcommand{\END}{{\CODE end}}
\newcommand{\RETURN}{{\CODE return}}
\newcommand{\WHILE}{{\CODE while}}

\begin{figure}[!t]
\begin{minipage}{2.2in}
{\fontsize{6}{7}
\begin{alltt}
{\FUNCT} {\color{red}{SWAY}}(optionSpace, evaluationFun,
              stop = sqrt(length(rows)))
              
{\INP}:  stop         : when to stop searching
        optionSpace  : dataset of possible options
        evaluationFun: run  model, get Table~\ref{evals} scores      
{\OUT}: selected node

selectionFun = all {\color{aoc} // See Table \ref{select}.}
half = {\color{red}HALF}(rows)
{\IF} count(half.lefts, half.rights) > stop
   {\IF} {\color{red}BEST\_LEFT}(half, evaluationFun) {\THEN}
      return {\color{red}SWAY}(half.left, evaluationFun, stop)
   {\ELSE} 
      return {\color{red}SWAY}(half.right, evaluationFun, stop)
{\ELSE}
    survivors = join(half.lefts, half.rights)
    {\RETURN} selectionFun(survivors)

\hrule 
{\FUNCT} {\color{red}{BEST\_LEFT}}(node, evaluationFun)

{\INP}:  node          :  node being evaluated
         evaluationFun :  method to evaluate an option                    
{\OUT}:  true if the left subtree is better than the right 

leftScore = evaluationFun(node.left)
rightScore = evaluationFun(node.right)
{\RETURN} dominates(leftScore, rightScore)  
\end{alltt}}
\end{minipage}\begin{minipage}{.7in}~\end{minipage}~\begin{minipage}{3in}
{\fontsize{6}{7}
\begin{alltt} \texttt
{\FUNCT} {\color{red}{HALF}}(rows,
              stop=sqrt(length(rows)),
              furthest=.95)

{\INP}: rows     : N rows, each with x independent 
                  and y dependent columns
        stop    : stopping criteria
        furthest: when looking for distant points, 
                  avoid outliers, don’t go
                  all the way
{\OUT}:  the rows, divided into two

tmp = any(rows)   {\color{aoc}// picked at random}
left = furthest row from tmp 
right = furthest row from left
{\FOR} row {\IN} rows {\DO} 
   a = dist(row, left)
   b = dist(row, right) 
   c = dist(left, right)
   row.x = (a^2 + c^2 - b^2) / (2c)  {\END}
lefts,rights = [], []
{\FOR} i,row {\IN} enumerate(sort(rows, key=x)) {\DO}
   {\IF}    i< length(rows)/2 
   {\THEN} push(lefts, row) 
   {\ELSE}  push(rights, row) 
   {\END} 
{\END}
{\RETURN} \{left=left, right=right, lefts=lefts, rights=rights\}

\end{alltt}}
\end{minipage} 
\caption{{\SWAY}, pseudo-code.}\label{sway}
\end{figure}

\subsection{SWAY} 

Chen et al.'s {\SWAY}~\cite{chen2018sampling} algorithm was discovered by accident
while  porting a JAVA version of a genetic algorithm  optimizer to PYTHON. Genetic algorithms use a mutator to explore different options.  Chen accidentally broke that mutator, then fixed it. Puzzlingly, after the fix,  the    resulting optimizations did not improve    (where ``improvement'' was measured by
the  Zitzler indicator).
On investigation Chen found that, given a large enough initial population,   mutation was not necessary. Standard genetic algorithms mutate
a population of 100 individuals for 100 generations~\cite{holland92}. Chen found he could optimize just as well using one generation of 10,000 randomly generated individuals (with no subsequent mutation or genetic cross-over).

The key to this approach was a simplistic greedy landscape analysis that   recursively clustered the candidates, using extreme examples. Informally, we   say that {\SWAY}
  draws a ``rope'' between two distant ``peaks'',  measures the height of the peaks,
then focuses on all the land nearer the higher ``peak''. More precisely, 
in  an approach inspired by 
Faloutsos et al.~\cite{faloutsos1995fastmap}, {\SWAY} evaluated two distant examples, then pruned   50\% of the data closest to the worst point 
\footnote{This can be done in linear time, as follows.
Pick any option $Z$ at random
then find the options $A,B$ most different to $Z,A$
(respectively).  Let $c={\mathit dist}(A,B)$ and let every   point have distances $a_i,b_i$ to $A,B$ respectively. Project
every point   onto a line between $A,B$ at a distance $x=(a_i^2 +c^2-b_i^2)/(2c)$ from $A$. Divided the data in two using the   median $x$ value. } (where ``worst'' was defined
by the Zitzler indicator described above).
When repeated recursively,
{\SWAY} finds   {\em best}=$\sqrt{N}$ best  options within $N$ examples after just  $2\log_2({N})$ evaluations
(i.e. one evaluation for each of the two distance examples found at each level of the recursion).

For more details on {\SWAY}, see
the pseudo code of Figure~\ref{sway}.

\subsection{{\IT}}\label{aboutit}

\begin{wrapfigure}{r}{2.2in}
\fontsize{6}{7}
\begin{alltt} 
{\FUNCT} {\color{red}{TREE}}(rows, stop=sqrt(N))

{\INP}:  rows : N rows, each with x independent 
               and y dependent columns    
        stop : stopping criteria
    
{\OUT}: balanced binary tree 

here= \{ rows=rows \}
{\IF} |rows| < stop {\THEN} {\RETURN} here {\END}
here.left, here.right, lefts, rights = {\color{red}HALF}(rows, top)
here.lefts = {\color{red}TREE}(lefts, stop)
here.right = {\color{red}TREE}(rights, stop)
{\RETURN} here
\end{alltt}
\caption{{\IT} builds a tree of all options (using {\color{red}{HALF}}, from Figure~\ref{sway}).}\label{tree}
\end{wrapfigure}
Conceptually, 
{\SWAY} builds a cluster tree of options but is hardwired to always select the split at the top of the current tree.
The central intuition of   {\IT} is that  better results can be  generated  by  looking around a little before committing to a particular jump.

~{\IT}'s tree construction algorithm
is the $\mathit{TREE}$ algorithm of 
Figure~\ref{tree}. Unlike the {\SWAY} code of Figure~\ref{sway}, we see that $\mathit{TREE}$     returns the entire tree of splits.
%After $\mathit{TREE}$, {\IT} applies 
%an algorithm called xPASS. In this part of the algorithm, 
{\IT}  then reflects on {\em every} sub-tree that divides $n_i$ examples
into two sub-trees of size $n_j,n_k$ (where $n_i=n_j+n_k$).

To say that another way, {\SWAY} is a local greedy search that prunes as soon as it can. {\IT}, on the other hand, is a more reflective algorithm that studies more of the total space.

For that reflection,
{\IT} uses entropy.
After all   numerics are discretized into  equal sized bins,   the entropy of a set of rows   is the sum of the entropy of the columns found in those rows\footnote{The entropy of a set of symbols is analogous to the standard deviation of  set of numbers 
(in that both are  a measure of how much we expect that distribution lies {\em away}
from the central tendency). Entropy can be thought of as the effort require to ``round up'' $1..i..n$ signals of probability $p_i$.
A binary chop can isolate a signal with effort $\log_2{p_i}$. For distribution $i$, the odds that we will need to utilize that effort
is $p_i$. Summing over the all the distributions, we say that entropy is     $-\sum_i^np_i\log_2{p_i}$.}.
 The  `best sub-tree'' with 
  the ``best seperation'' is the one with greatest difference in the entropy between the root of the sub-tree and its two children. To avoid trivial cases (near the leaves), this entropy value is weighted with the number of rows $n_i$ found in the root $i$ and   two sub-trees $j,k$:
  \begin{equation}\label{choice}
   \begin{array}{rcl}
  w_1&=& n_i\times \mathit{entropy}_i \\
    w_2&=& n_j\times \mathit{entropy}_j + 
               n_k\times \mathit{entropy}_k \\
   \mathit{best}&=& w_1-w_2
    \end{array} \end{equation}

Once a  ``best sub-tree'' with 
  the ``best seperation''  is found, 
  our sub-routine   
  ``xPASS''  evaluates two extreme points in that split and prunes the sub-tree containing
the worst options.  xPASS
    queries sub-trees that
  \bi
  \item Have not been queried before;
  \item Have different values in the 
$n_j$ and $n_k$ rows
\item Have high variability; i.e. has the largest entropies.
\ei
Once the best sub-tree is found, xPASS finds and deletes   rows of the ``worst'' sub-tree. 
After that, it loops   looking for other sub-trees to probe
(stopping when the tree is smaller than $\sqrt{N}$). 

In summary, {\IT} uses the nested functions of  Equation~\ref{sneak}:

 \begin{equation}\label{sneak} % aboutit
\mathit{SNEAK}(\mathit{rows})=
 \underbrace{\mathit{SELECT}}_{\mathit{Table~\ref{select}}}(\;
  \mathit{yPASS}(
    \underbrace{\mathit{xPASS}}_{\mathit{see \;section\; 3.3}}(
        \mathit{TREE}(
           \mathit{rows})))
       \end{equation}

The ``yPASS'' sub-routine is where we configure, then execute a pair  of  machine 
learning models (in this case here, random forests)
and only the best half of each pair is collected for SELECT.
Given all the work done on the right-hand-side of yPASS,  we need to run far fewer evaluations than
other approaches  (for exact statistics on that, see Table~\ref{tab:globmedian}).  

{\IT} outputs around $\sqrt{N}$ possible configurations. Hence, some a {\em SELECT} function must be applied in order to select the best among these final configurations. 
For the first two research questions of this paper, we apply the following procedure:
 (a)~pass the $\sqrt{N}$ items returned from  yPASS to a second call to SWAY;
 (b)~evaluate all the 
  $\sqrt{\sqrt{N}}$ items selected this manner; (c)~return the best one (as judged by the   $\mathit{D2H}_s$ measure of Table~\ref{evals}).
In
{\bf RQ3} of this paper explore variants to that procedure.

It has not escaped our attention that   {\IT} need not run fully automatically. Rather, in Equation~\ref{choice}, we could ask  for human judgment to decide which sub-tree   to be prune.
As part of his Ph.D. dissertation, our first author is exploring this interactive version of {\IT}. That work
assumes that the
extra constraints learned from humans   is   most useful for the optimization of  reasoning over very large data sets (as opposed to  the tiny data sets explored here).

\section{Experimental Methods}\label{methods}

This paper uses software project health data from Xia et al. (obtained from their GitHub repository) to comparatively assess the hyperparameter optimizers of \S\ref{hpola} with the landscape analytics of {\IT}.
Several of our algorithms have some stochastic components. 
Hence  following the advice of Arcuri and Briand~\cite{arcuri2011practical}, our experimental rig repeats the following procedure 20 times, each time using different random number seeds.

At MSR'22, Majumder et al.~\cite{Majumder22} reported twelve clusters of projects within the projects explored by Xia et al. 
This paper applies our methods to the projects nearest to those twelve clusters.
Table~\ref{tab:features} shows statistics for a dozen features collected from those 12 projects.
Data was originally collected from these 12 projects in one-month increments. These projects exist for different lengths of time, 
\begin{wraptable}{r}{3.5in}
\caption{Summary of project health datasets used in this study.
Median is the 50th percentile and IQR (short for ``inter-quartile range'') is the (75-25)th percentile.}
\footnotesize
\begin{center}
\begin{tabular}{l|cccc}
                        & \textbf{Min} & \textbf{Max} & \textbf{Median} & \textbf{IQR} \\ \hline
\textbf{Commits}        & 0            & 919          & 22              & 69           \\
\textbf{Contributors}   & 0            & 29           & 2               & 6            \\
\textbf{Open PRs}       & 0            & 82           & 1               & 5            \\
\textbf{Closed PRs}     & 0            & 17           & 0               & 1            \\   
\textbf{Merged PRs}     & 0            & 77           & 0               & 3            \\
\textbf{PR comments}    & 0            & 1098         & 0               & 7            \\
\textbf{Open issues}    & 0            & 137          & 1               & 4            \\
\textbf{Closed issues}  & 0            & 385          & 8               & 31           \\
\textbf{Issue comments} & 0            & 1553         & 17              & 96           \\
\textbf{Stars}          & 0            & 2543         & 25              & 54           \\
\textbf{Forks}          & 0            & 379          & 10              & 12           \\
\textbf{Watchers}       & 0            & 15           & 0               & 1        
\end{tabular}
\end{center}
\label{tab:features}
\vspace{-10pt}
\end{wraptable}
specifically 32 to 63 months.
This meant our 12 projects had between 32 to 63 rows.
We generated  predictions 12 months into the future.
To do so, we used all rows from 12 months prior to predicting for the 12 months.
For example, if a project had  40 months of data, then we
predicted for months 29,30,31, ..., 40 using
just the data found in the first 17, 18, 19, ..., 28 rows
of data (respectively). 
For the target   of this learning, we used Table~\ref{tbl:survey} to select for targets   that would most prompt core developers to take action within their projects; i.e.:

\be
    \item {\em Indicator1}= number of commits;
    \item {\em Indicator2}= number of closed issues;
    \item {\em Indicator3}= number of closed pull requests.
\ee
Then, for each indicator, we built predictors using  that indicator as to the $Y$ features, and every other feature as independent $X$ features.  
This means that within our 20 repeats, we:
\bi
    \item Looped three  times (once for each indicator)
    \item Looping other times (once for each data set) for
         \bi
            \item The seven hyperparameter optimizers of \S\ref{tab:paramGrids}
            \item The SELECT operator options explored in {\bf RQ3};
             \item Plus one   call  with   off-the-shelf {\em default} parameters.
         \ei
\ei
When collection performance scores, we report median (50th percentile) and IQR (75-25)th percentile. We  use these non-parametric
measures since 
  estimation is a domain with many large outliers-- exactly the sort of space where measurements of ``mean'' can misrepresent the data.

For the above $20*3*12*12=8,640$ runs (except for the {\em default} runs), we tuned the hyperparameters of a random forest. 
This learner was selected since a wide range of
software analytics researchers have reported
that random forests are very useful for this kind of estimation~\cite{weber2014makes, kikas2016using, mustapha2019investigating, banimustafa2018predicting, xia2018hyperparameter}. 

Table~\ref{tab:paramGrids} shows the space of hyperparameters we explored. 
Four of these hyperparameters were chosen since they were also used by Xia et al. while the fifth (number of estimators) is important for random forests (since it sets the size of the forest). 

Different optimizers treated the ranges of
Table~\ref{tab:paramGrids} in different ways. For example,
DE extrapolating any value at all between the min and max values. As to {\IT}, that algorithm selected at random from   used 20 steps between min and max.
Also, for continuous parameters (min impurity), 
{\IT} did a more fine-grained approach and used 40 steps. 
The cross product of Table~\ref{tab:paramGrids} returns 984,000 possible model configurations.  
From this space, we randomly selected 10,000 models, ran and cached the results. 
Each tuning algorithm was then scored by how few of those evaluation results they used in their reasoning. Without that caching,
our experiments would take weeks to terminate but
with caching, on a 16-core machine~\footnote{Windows 10 machine with an Intel i7-11700K, 64GB of DDR4 Ram and an Nvidia RTX 3080}, our experiments terminated in under six hours.

% XXX 20 times randm seeds

% \begin{table}[!t]
% \caption{Parameter Grid for generating {\IT}'s option space. (Note that criterion is a discrete parameter)}
% \footnotesize
% \begin{tabular}{l|lll}
% \textbf{Parameter}               & \textbf{Min}         & \textbf{Max}         & \textbf{Step}        \\ \hline
% \textbf{n\_estimators}           & 10                   & 200                  & 10                   \\
% \textbf{min\_sample\_leaves}     & 1                    & 20                   & 1                    \\
% \textbf{min\_impurity\_decrease} & 0                    & 10                   & 0.25                 \\
% \textbf{max\_depth}              & 1                    & 20                   & 1                    \\ 
% \textbf{criterion}               & \multicolumn{3}{c} {one of: [squared, absolute, poisson]}
% \end{tabular}
% \label{tab:paramGrid}
% \vspace{-10pt}
% \end{table}

\subsection{Algorithms}

The {\IT} system is implemented in Python 3.8.
That code is made available on-line at \github.
In order to run most of our baseline algorithms, we have used the NUE framework~\footnote{\url{https://github.com/lyonva/Nue}}.
The NUE framework contains the implementation of most of the state-of-the-art hyperparameter optimization algorithms as an extension to scikit-learn~\cite{sklearn_api}. 
Through this framework, we have executed our runs for Random Search, {\SWAY}, Differential Evolution, Grid Search and Flash.

In order to use HYPEROPT and OPTUNA, we have made use of the implementations made publicly available through pip. 
Which can be installed via ``pip3 install hyperopt'' and ``pip3 install optuna''.
To run RF with its default parameters we have simply trained our learner on all target features using standard scikit-learn settings.

\subsection{Statistical Methods}\label{stats}

We present below two kinds of results. {\em Firstly}, we offer the
standard summary statistics:
\bi 
\item Median   (50th percentile) results of our methods' performance across all datasets. 
\item  Inter-quartile range results (IQR$=75\mathit{th}-25\mathit{th}$ percentile) of the same data.
\ei
{\em Secondly},
in order to rank these methods, and decide if any method is better we use a combination of the Friedman test with the Nemenyi Post-Hoc test. 
This methodology follows from suggestions from Dem{\v{s}}ar et al.~\cite{demvsar2006statistical}, Herbold et al~\cite{herbold2017comments, herbold2018correction} and Xia et al.~\cite{xia2018hyperparameter}.

The Friedman test is a non-parametric test that determines whether or not there is a statistically significant difference between populations~\cite{friedman1940comparison}. 
As such when comparing results for our treatments on a particular metric predicting a particular indicator we first check for the presence of a statistical significant difference. 
For our case, we run this test, and if the corresponding p-value is larger than $0.05$ we consider that there is no significant difference between the methods for this indicator and metric.
If the p-value is, however, lower than $0.05$, we then run the Post-Hoc test~\cite{nemenyi1963distribution} to determine the groupings of treatments that are significantly different. 
This test uses a measure of Critical Distances between average ranks to define significantly different populations. 
% As a threshold for separating these groups, we also set the significance level to $0.05$.

\begin{table}[b!]
    
      \tiny   \begin{tabular}[t]{lccccccccccccccc}
\multicolumn{1}{l}{~} &  \rotatebox{60}{commits}              & \rotatebox{60}{closedPRs}      & \rotatebox{60}{closedIssues}  && 
                        \rotatebox{60}{commits}              & \rotatebox{60}{closedPRs}      & \rotatebox{60}{closedIssues}  &&  
                        \rotatebox{60}{commits}              & \rotatebox{60}{closedPRs}      & \rotatebox{60}{closedIssues}  &&  
                        \rotatebox{60}{commits}              & \rotatebox{60}{closedPRs}      & \rotatebox{60}{closedIssues}  \\                                                
\multicolumn{1}{l}{{\SWAY}}    &{\DARK}\textcolor{white}{52} & {\DARK}\textcolor{white}{52}   & {\MID}52                     &&
                             {\DARK}\textcolor{white}{48}    & {\DARK}\textcolor{white}{42}   & {\DARK}\textcolor{white}{39} &&
                             {\DARK}\textcolor{white}{26}    & {\DARK}\textcolor{white}{34}   & {\DARK}\textcolor{white}{23} &&
                             {\DARK}\textcolor{white}{45\%}  & {\DARK}\textcolor{white}{45\%} & {\DARK}\textcolor{white}{45\%}                    \\               
\multicolumn{1}{l}{Default} &{\MID}70                       & {\MID}100                      & {\LIGHT}66                    &&
                             {\MID}23                        & {\MID}14                       & {\MID}32                      &&
                             {\LIGHT}14                      & {\MID}-6                       & {\LIGHT}-5                    &&
                             {\MID}45\%                      & {\MID}50\%                     & {\MID}59\%                    \\ 
\multicolumn{1}{l}{RS}      &{\DARK}\textcolor{white}{51}                       & {\DARK}\textcolor{white}{29}                       & {\MID}43                    &&
                             {\DARK}\textcolor{white}{48}                        & {\DARK}\textcolor{white}{60}                       & {\DARK}\textcolor{white}{42}                      &&
                             {\DARK}\textcolor{white}{26}                        & {\DARK}\textcolor{white}{36}                       & {\DARK}\textcolor{white}{23}                      &&
                             {\DARK}\textcolor{white}{27}\%                      & {\DARK}\textcolor{white}{18\%} & {\DARK}\textcolor{white}{27}\%                    \\
\multicolumn{1}{l}{DE}      &{\MID}78                       & {\MID}80                       & {\LIGHT}50                    &&
                             {\MID}23                        & {\LIGHT}7                      & {\MID}36                      &&
                             {\MID}15                        & {\MID}-2                       & {\MID}6                       &&
                             {\MID}32\%                      & {\MID}54\%                     & {\MID}50\%                    \\
\multicolumn{1}{l}{GS}      &{\MID}40                       & {\MID}34                       & {\LIGHT}57                    &&
                             {\MID}53                        & {\MID}59                       & {\MID}32                      &&
                             {\DARK}\textcolor{white}{32}    & {\MID}33                       & {\MID}16                      &&
                             {\MID}32\%                      & {\MID}50\%                     & {\MID}36\%                    \\
\multicolumn{1}{l}{FLASH}   &{\MID}77                       & {\LIGHT}94                     & {\LIGHT}66                    &&
                             {\MID}17                        & {\LIGHT}0                      & {\MID}28                      &&
                             {\MID}19                        & {\LIGHT}-10                    & {\MID}15                      &&
                             {\MID}36\%                      & {\LIGHT}72\%                   & {\MID}54\%                    \\
\multicolumn{1}{l}{HYPEROPT}&{\DARK}\textcolor{white}{55}    & {\DARK}\textcolor{white}{33}   & {\DARK}\textcolor{white}{35}  &&
                             {\LIGHT}5                       & {\LIGHT} 7                     & {\LIGHT}7                     &&
                             {\LIGHT}-1                      & {\LIGHT} 0                       & {\LIGHT}0                       &&
                             {\LIGHT}77\%                    & {\DARK}45\%                     & {\LIGHT}86\%                  \\
\multicolumn{1}{l}{OPTUNA}  &{\LIGHT}149                    & {\LIGHT} 119                   & {\LIGHT}61                    &&
                             {\LIGHT}2                       & {\LIGHT} 0                     & {\LIGHT}4                     &&
                             {\LIGHT}0                       & {\MID} 3                       & {\MID}0                       &&
                             {\LIGHT}99\%                    & {\LIGHT}77\%                   & {\LIGHT}86\%                  \\  
& \multicolumn{3}{c} {(a) Median MRE  }    && \multicolumn{3}{c} {(b) Median Pred40  }   && \multicolumn{3}{c} {(c)  Median SA  }   && \multicolumn{3}{c} {(d) Median D2H  }\\ 
~\\
\multicolumn{16}{c}{\small{{\bf Table \ref{tab:RQ1}.a: Median results (over 20 runs).}}}\\
  &  \rotatebox{60}{commits}              & \rotatebox{60}{closedPRs}      & \rotatebox{60}{closedIssues}  && 
                        \rotatebox{60}{commits}              & \rotatebox{60}{closedPRs}      & \rotatebox{60}{closedIssues}  &&  
                        \rotatebox{60}{commits}              & \rotatebox{60}{closedPRs}      & \rotatebox{60}{closedIssues}  &&  
                        \rotatebox{60}{commits}              & \rotatebox{60}{closedPRs}      & \rotatebox{60}{closedIssues}  \\  
                        
& \multicolumn{3}{c} {   (lower values are better)}    && \multicolumn{3}{c} {   (higher values are  better)}   && \multicolumn{3}{c} {  (higher values are  better)}   && \multicolumn{3}{c} {   (lower values are  better)}\\   

\multicolumn{1}{l}{{\SWAY}}    &{\LIGHT}72                  & {\LIGHT}100                    & {\LIGHT}50                    &&
                             {\LIGHT}55                      & {\LIGHT}68                     & {\LIGHT}44                    &&
                             {\LIGHT}17                      & {\LIGHT}85                     & {\LIGHT}8                     &&
                             {\LIGHT}20\%                    & {\LIGHT}20\%                   & {\LIGHT}41\%                  \\ 
\multicolumn{1}{l}{Default} &{\LIGHT}82                     & {\LIGHT}0                      & {\LIGHT}98                    &&
                             {\LIGHT}41                      & {\LIGHT}31                     & {\LIGHT}49                    &&
                             {\LIGHT}81                      & {\LIGHT}15                     & {\LIGHT}52                    &&
                             {\LIGHT}36\%                    & {\LIGHT}83\%                   & {\LIGHT}47\%                  \\  
\multicolumn{1}{l}{RS}      &{\LIGHT}69                     & {\LIGHT}99                     & {\LIGHT}50                    &&
                             {\LIGHT}41                      & {\LIGHT}60                     & {\LIGHT}44                    &&
                             {\LIGHT}15                      & {\LIGHT}66                     & {\LIGHT}22                    &&
                             {\LIGHT}29\%                    & {\LIGHT}20\%                   & {\LIGHT}47\%                  \\
\multicolumn{1}{l}{DE}      &{\LIGHT}104                    & {\LIGHT}50                     & {\LIGHT}112                   &&
                             {\LIGHT}37                      & {\LIGHT}27                     & {\LIGHT}40                    &&
                             {\LIGHT}68                      & {\LIGHT}16                     & {\LIGHT}45                    &&
                             {\LIGHT}50\%                    & {\LIGHT}20\%                   & {\LIGHT}36\%                  \\
\multicolumn{1}{l}{GS}      &{\LIGHT}49                     & {\LIGHT}100                    & {\LIGHT}48                    &&
                             {\LIGHT}43                      & {\LIGHT}62                     & {\LIGHT}53                    &&
                             {\LIGHT}15                      & {\LIGHT}84                     & {\LIGHT}24                    &&
                             {\LIGHT}32\%                    & {\LIGHT}54\%                   & {\LIGHT}29\%                  \\
\multicolumn{1}{l}{FLASH}   &{\LIGHT}68                     & {\LIGHT}41                     & {\LIGHT}87                    &&
                             {\LIGHT}38                      & {\LIGHT}15                     & {\LIGHT}37                    &&
                             {\LIGHT}62                      & {\LIGHT}15                     & {\LIGHT}45                    &&
                             {\LIGHT}63\%                    & {\LIGHT}14\%                   & {\LIGHT}50\%                  \\
\multicolumn{1}{l}{HYPEROPT}&{\LIGHT}131                    & {\LIGHT}76                     & {\LIGHT}76                    &&
                             {\LIGHT}4                       & {\LIGHT}6                      & {\LIGHT}6                     &&
                             {\LIGHT}7                       & {\LIGHT}18                     & {\LIGHT}15                    &&
                             {\LIGHT}36\%                    & {\LIGHT}50\%                   & {\LIGHT}20\%                  \\
\multicolumn{1}{l}{OPTUNA}  &{\LIGHT}157                    & {\LIGHT}52                     & {\LIGHT}76                    &&
                             {\LIGHT}4                       & {\LIGHT}3                      & {\LIGHT}5                     &&
                             {\LIGHT}15                      & {\LIGHT}24                     & {\LIGHT}30                    &&
                             {\LIGHT}14\%                    & {\LIGHT}38\%                   & {\LIGHT}36\%                  \\ 
& \multicolumn{3}{c} {(e) MRE IQR}    && \multicolumn{3}{c} {(f) Pred40 IQR}   &&\multicolumn{3}{c} {(g) SA IQR}      && \multicolumn{3}{c} {(h) D2H IQR} \\
~\\
\multicolumn{16}{c}{\small{{\bf Table \ref{tab:RQ1}.b: IQR results (over 20 runs).}}}\\~\\

        \end{tabular}
        \caption{ RQ1 results.  Median and IQR results (over 20 runs) for the 
        Table~\ref{evals} measures. RS= random search. PR= pull requests.
        Median = 50th  percentile.     IQR =(75-25)th percentile. {\em Darker} cells show {\em better} results. Here ``better'' is computed relative to other results in the same column, using the statistical methods of \S\ref{stats}.  In this table {\em lower} values for MRE and D2H are {\em better}
    while {\em higher} values for SA and Pred40 are {\em better}. The negative values
    seen in  median SA are not errors-- that measure can   fall below zero. 
    }
        \label{tab:RQ1}
\end{table}

\section{Results}\label{results}

This section offers answers to the research questions listed in the introduction. 
These questions are answered using the result presented in 
Tables~\ref{tab:RQ1},\ref{tab:RQ2} and \ref{tab:RQ3}. 
All these tables show results after trying to predict for
project health indicators using data sets containing around 60 rows and 14 columns from Table~\ref{healthind}.
Our result tables show
 the median (50th percentile) and IQR (75th - 25th percentile) over 20 runs. 
The cells of   Tables~\ref{tab:RQ1},\ref{tab:RQ2} and \ref{tab:RQ3}  are colored according to the results of  a per-column statistical analysis
(of the form defined in \S\ref{stats}). Specifically:  the darker the cell, the more than treatment out-performs other
treatments in that column (so the best results are shown in the {\em darkest} cell).

\subsection{RQ1: Does small sample size damage our ability to perform project health prediction?}

Recall from the introduction that  this is our {\em baseline} question. 
The   motivation of this paper is that  learning from small
sample sizes   can lead to errors in project health estimation. 
To document that claim, Table~\ref{tab:RQ1} shows results where standard methods try to predict for
project health indicators.

There are many noteworthy features of Table~\ref{tab:RQ1}:
\bi
\item 
Our introduction defined acceptable errors for this kind of estimation at 30\% ot 40\% (measured as MRE;
i.e. {\em abs(actual -
predicted)/ actual)}). Many of the results in Table~\ref{tab:RQ1} exceed that limit.  
\item Many   methods perform worse than random search (indicated by ``RS'' in that table).    \ei
Why might random search beat supposedly more sophisticated methods?  
Consider the   OPTUNA results in Table~\ref{tab:RQ1}.
This state-of-the-art algorithm  has a well-documented superior performance on much larger data sets. However,
on the smaller data sets used here, it
was unable to reach good optimization results. 
To explain this, recall from the above that OPTUNA uses MOTPE
(the  optimization of multiple performance metrics at once by analyzing the results of the current models against the currently observed frontier of solutions via using a function that calculates the expected hypervolume improvement at each step of its optimization).
We conjecture   that the hypervolume calculation used in    MOTPE is not suited for such small datasets as the ones we are working with here. (i.e. 20 to 40 data points). 

Also note that the performance of Hyperopt that performs well on MRE but no so well in the other goals.
Hyperopt only considers single goal problems and, in this experiment, we focused on MRE-- so it is hardly surprising that it did well on one   measure, and not the others.

In summary, our answer to {\bf RQ1} is that:
\begin{RQ}
 Standard methods cannot handle the small data sets seen in our domain.    
\end{RQ}

\begin{table}[b!]
  
      \tiny   \begin{tabular}[t]{lccccccccccccccc}
\multicolumn{1}{l}{\underline{Algorithm}} &  \rotatebox{60}{commits}              & \rotatebox{60}{closedPRs}      & \rotatebox{60}{closedIssues}  && 
                        \rotatebox{60}{commits}              & \rotatebox{60}{closedPRs}      & \rotatebox{60}{closedIssues}  &&  
                        \rotatebox{60}{commits}              & \rotatebox{60}{closedPRs}      & \rotatebox{60}{closedIssues}  &&  
                        \rotatebox{60}{commits}              & \rotatebox{60}{closedPRs}      & \rotatebox{60}{closedIssues}  \\                                                
\multicolumn{1}{l}{{\IT}}   &{\MID}47                  & {\DARK}\textcolor{white}{0}    & {\MID}33                      &&
                             {\DARK}\textcolor{white}{43}    & {\DARK}\textcolor{white}{93}   & {\DARK}\textcolor{white}{60}  &&
                             {\DARK}\textcolor{white}{33}    & {\DARK}\textcolor{white}{47}   & {\DARK}\textcolor{white}{19}  &&
                             {\DARK}\textcolor{white}{5\%}   & {\DARK}\textcolor{white}{5\%}  & {\DARK}\textcolor{white}{5\%} \\
\multicolumn{1}{l}{{\SWAY}}    &{\MID}52                    & {\MID}52                       & {\LIGHT}52                    &&
                             {\MID}48                        & {\MID}42                       & {\MID}39                      &&
                             {\MID}26                        & {\MID}34                       & {\MID}23                      &&
                             {\MID}45\%                      & {\MID}45\%                     & {\MID}45\%                    \\
\multicolumn{1}{l}{RS}      &{\MID}51                       & {\MID}29                       & {\LIGHT}43                    &&
                             {\MID}48                        & {\MID}60                       & {\MID}42                      &&
                             {\MID}26                        & {\MID}36                       & {\MID}23                      &&
                             {\MID}27\%                      & {\DARK}\textcolor{white}{18\%} & {\MID}27\%                    \\
\multicolumn{1}{l}{HYPEROPT}&{\MID}55                       & {\MID}33                       & {\MID}35                      &&
                             {\LIGHT}5                       & {\LIGHT} 7                     & {\LIGHT}7                     &&
                             {\LIGHT}-1                      & {\MID} 0                       & {\MID}0                       &&
                             {\LIGHT}77\%                    & {\MID}45\%                     & {\LIGHT}86\%                  \\
& \multicolumn{3}{c} {(a) Median MRE  }    && \multicolumn{3}{c} {(b) Median Pred40  }   && \multicolumn{3}{c} {(c)  Median SA  }   && \multicolumn{3}{c} {(d) Median D2H  }\\ 
~\\
\multicolumn{16}{c}{\small{{\bf Table \ref{tab:RQ2}.a: Median results (over 20 runs).}}}\\
  &  \rotatebox{60}{commits}              & \rotatebox{60}{closedPRs}      & \rotatebox{60}{closedIssues}  && 
                        \rotatebox{60}{commits}              & \rotatebox{60}{closedPRs}      & \rotatebox{60}{closedIssues}  &&  
                        \rotatebox{60}{commits}              & \rotatebox{60}{closedPRs}      & \rotatebox{60}{closedIssues}  &&  
                        \rotatebox{60}{commits}              & \rotatebox{60}{closedPRs}      & \rotatebox{60}{closedIssues}  \\  
                        
& \multicolumn{3}{c} {   (lower values are better)}    && \multicolumn{3}{c} {   (higher values are  better)}   && \multicolumn{3}{c} {  (higher values are  better)}   && \multicolumn{3}{c} {   (lower values are  better)}\\

\multicolumn{1}{l}{{\IT}}   &{\LIGHT}85                & {\LIGHT}0                      & {\LIGHT}69                    &&
                             {\LIGHT}56                      & {\LIGHT}45                     & {\LIGHT}60                    &&
                             {\LIGHT}52                      & {\LIGHT}73                     & {\LIGHT}30                    &&
                             {\LIGHT}41\%                    & {\LIGHT}9\%                    & {\LIGHT}11\%                  \\
\multicolumn{1}{l}{{\SWAY}}    &{\LIGHT}72                  & {\LIGHT}100                    & {\LIGHT}50                    &&
                             {\LIGHT}55                      & {\LIGHT}68                     & {\LIGHT}44                    &&
                             {\LIGHT}17                      & {\LIGHT}85                     & {\LIGHT}8                     &&
                             {\LIGHT}20\%                    & {\LIGHT}20\%                   & {\LIGHT}41\%                  \\
\multicolumn{1}{l}{RS}      &{\LIGHT}69                     & {\LIGHT}99                     & {\LIGHT}50                    &&
                             {\LIGHT}41                      & {\LIGHT}60                     & {\LIGHT}44                    &&
                             {\LIGHT}15                      & {\LIGHT}66                     & {\LIGHT}22                    &&
                             {\LIGHT}29\%                    & {\LIGHT}20\%                   & {\LIGHT}47\%                  \\
\multicolumn{1}{l}{HYPEROPT}&{\LIGHT}131                    & {\LIGHT}76                     & {\LIGHT}76                    &&
                             {\LIGHT}4                       & {\LIGHT}6                      & {\LIGHT}6                     &&
                             {\LIGHT}7                       & {\LIGHT}18                     & {\LIGHT}15                    &&
                             {\LIGHT}36\%                    & {\LIGHT}50\%                   & {\LIGHT}20\%                  \\
& \multicolumn{3}{c} {(e) MRE IQR}    && \multicolumn{3}{c} {(f) Pred40 IQR}   &&\multicolumn{3}{c} {(g) SA IQR}      && \multicolumn{3}{c} {(h) D2H IQR} \\
~\\
\multicolumn{16}{c}{\small{{\bf Table \ref{tab:RQ2}.b: IQR results (over 20 runs).}}}\\~\\

        \end{tabular}
          \caption{RQ2 results.  Same format at Table~\ref{tab:RQ1}; i.e. 
     {\em Darker} cells show {\em better} results.  
    }
        \label{tab:RQ2}
\end{table}
\begin{table}[!b]
\footnotesize
\begin{tabular}{cp{3.3in}l}
\rowcolor{black!50}& & \color{white}{\#evaluations ($E$)}\\\hline
{\bf ANY} & Evaluate and return any  one  of the {\em best} items.  & $2\log_2({N})+1$\\ 
 
\rowcolor{gray!15}{\bf SANY} & Pass {\em best} to a second round of    {\SWAY} to return $\sqrt{\sqrt{N}}$ items.
Return any item from this set.
&
  $2\log_2({N})+2\log_2({\sqrt{N}})$ \\
  
{\bf ALL} & Evaluate  {\em best};  return the one 
with smallest  D2H (see Table \ref{evals}).
&   $2\log_2({N}) + \sqrt{N}$.\\
 
\rowcolor{gray!15}{\bf SALL}&   Pass {\em best} to a second round of {\SWAY};
apply {\bf ALL} to the   $\sqrt{\sqrt{N}}$ items found in this way. & 
$2\log_2({N})+2\log_2({\sqrt{N}}) +
\sqrt{\sqrt{N}}$.
\end{tabular}

\caption{Four   SELECT operators for
finding final items to report to user. All these methods require
an initial $2\log_2({N})$ evaluations to complete the method described in \S\ref{sneak}. }\label{select}
\end{table}

\subsection{RQ2: Can landscape analysis mitigate for small sample sizes?}
Here,  we compared the performance of  {\IT} to the better performing methods  from {\bf RQ1}.  
Table~\ref{tab:RQ2} compares the performance of {\IT} to randoms search (RS), SWAY and HYPEROPT (we use HYPEROPT since regardless of its performance overall,
it did perform well on MRE).

The most noteworthy feature of Table~\ref{tab:RQ2}  is that across all measures, {\IT} performed as well, or better, than anything else. It is also encouraging to see that {\IT} performs better than random search (since, otherwise, this whole paper would be a needless over-elaboration of something that could implemented via  a much simpler random process).  We conjecture that {\IT} works so well since it finds the most informative
regions of the hyperparameters, then jumps to those regions. Other methods (that do not reflect over the
landscape) can waste time exploring less informative options.

In summary, our answer to {\bf RQ2} is that:
\begin{RQ}
At least for project health estimation,
landscape analysis tools like {\IT} can  mitigate for small sample sizes.   
\end{RQ}

\subsection{RQ3: How much landscape analysis is enough?}

To explain {\bf RQ3}, we   return to \S\ref{aboutit}
and the discussion around Equation~\ref{sneak}. There, we noted that {\IT} can return multiple
solutions so some SELECT operator must reduce those many  down to one. 
Table~\ref{select} offers several SELECT operators (and note that for {\bf RQ2}, we used the ``SALL'' option seen in row four
of that table).
 For example, the ``ANY'' option (seen in row one of Table~\ref{select})
makes the standard landscape analysis assumption (discussed in \S\ref{latics}) that all the solutions can be fully enumerated before being studied. Other methods (e.g. ``SANY'' and ``SALL'') call our tree generation routines to 
(1)~find some tiny subset before (2)~evaluating just that tiny set.

(Aside: 
We make no claim that Table~\ref{select} lists the space of all possible SELECT methods. 
That said, this set is at least interesting since it  can still arrive some very low error solutions.)

\begin{table}[b!]
  
      \tiny   \begin{tabular}[t]{lccccccccccccccc}
\multicolumn{1}{l}{\underline{Algorithm}} &  \rotatebox{60}{commits}              & \rotatebox{60}{closedPRs}      & \rotatebox{60}{closedIssues}  && 
                        \rotatebox{60}{commits}              & \rotatebox{60}{closedPRs}      & \rotatebox{60}{closedIssues}  &&  
                        \rotatebox{60}{commits}              & \rotatebox{60}{closedPRs}      & \rotatebox{60}{closedIssues}  &&  
                        \rotatebox{60}{commits}              & \rotatebox{60}{closedPRs}      & \rotatebox{60}{closedIssues}  \\                                                
\multicolumn{1}{l}{{\IT}+all} &{\DARK}\textcolor{white}{38} & {\DARK}\textcolor{white}{0}    & {\DARK}\textcolor{white}{26}  && 
                             {\DARK}\textcolor{white}{52}    & {\DARK}\textcolor{white}{93}   & {\DARK}\textcolor{white}{66}  &&
                             {\DARK}\textcolor{white}{36}    & {\DARK}\textcolor{white}{46}   & {\DARK}\textcolor{white}{27}  &&
                             {\MID}54\%                      & {\DARK}\textcolor{white}{18\%} & {\MID}27\%                    \\
\multicolumn{1}{l}{{\IT}+Sall}   &{\MID}47                  & {\DARK}\textcolor{white}{0}    & {\MID}33                      &&
                             {\DARK}\textcolor{white}{43}    & {\DARK}\textcolor{white}{93}   & {\DARK}\textcolor{white}{60}  &&
                             {\DARK}\textcolor{white}{33}    & {\DARK}\textcolor{white}{47}   & {\DARK}\textcolor{white}{19}  &&
                             {\DARK}\textcolor{white}{5\%}   & {\DARK}\textcolor{white}{5\%}  & {\DARK}\textcolor{white}{5\%} \\  
\multicolumn{1}{l}{{\IT}+Sany} &{\LIGHT}120                 & {\LIGHT}117                    & {\LIGHT}70                    &&
                             {\LIGHT}16                      & {\LIGHT}0                      & {\MID}24                      &&
                             {\LIGHT}8                       & {\MID}-5                       & {\MID}6                       &&
                             {\LIGHT}63\%                    & {\LIGHT}77\%                   & {\MID}41\%                    \\
\multicolumn{1}{l}{{\IT}+any} &{\LIGHT}120                  & {\LIGHT} 120                   & {\LIGHT}72                    &&
                             {\LIGHT}8                       & {\LIGHT}0                      & {\MID}24                      &&
                             {\LIGHT}2                       & {\MID}-5                       & {\LIGHT}6                     &&
                             {\LIGHT}72\%                    & {\LIGHT}81\%                   & {\MID}54\%                    \\
& \multicolumn{3}{c} {(a) Median MRE  }    && \multicolumn{3}{c} {(b) Median Pred40  }   && \multicolumn{3}{c} {(c)  Median SA  }   && \multicolumn{3}{c} {(d) Median D2H  }\\ 
~\\
\multicolumn{16}{c}{\small{{\bf Table \ref{tab:RQ3}.a: Median results (over 20 runs).}}}\\
  &  \rotatebox{60}{commits}              & \rotatebox{60}{closedPRs}      & \rotatebox{60}{closedIssues}  && 
                        \rotatebox{60}{commits}              & \rotatebox{60}{closedPRs}      & \rotatebox{60}{closedIssues}  &&  
                        \rotatebox{60}{commits}              & \rotatebox{60}{closedPRs}      & \rotatebox{60}{closedIssues}  &&  
                        \rotatebox{60}{commits}              & \rotatebox{60}{closedPRs}      & \rotatebox{60}{closedIssues}  \\  
                        
& \multicolumn{3}{c} {   (lower values are better)}    && \multicolumn{3}{c} {   (higher values are  better)}   && \multicolumn{3}{c} {  (higher values are  better)}   && \multicolumn{3}{c} {   (lower values are  better)}\\

\multicolumn{1}{l}{{\IT}+all} &{\LIGHT}59                   & {\LIGHT}0                      & {\LIGHT}71                    && 
                             {\LIGHT}51                      & {\LIGHT}45                     & {\LIGHT}48                    &&
                             {\LIGHT}42                      & {\LIGHT}69                     & {\LIGHT}30                    &&
                             {\LIGHT}47\%                    & {\LIGHT}5\%                    & {\LIGHT}25\%                  \\
\multicolumn{1}{l}{{\IT}+Sall}   &{\LIGHT}85                & {\LIGHT}0                      & {\LIGHT}69                    &&
                             {\LIGHT}56                      & {\LIGHT}45                     & {\LIGHT}60                    &&
                             {\LIGHT}52                      & {\LIGHT}73                     & {\LIGHT}30                    &&
                             {\LIGHT}41\%                    & {\LIGHT}9\%                    & {\LIGHT}11\%                  \\
\multicolumn{1}{l}{{\IT}+Sany} &{\LIGHT}127                 & {\LIGHT}24                     & {\LIGHT}88                    &&
                             {\LIGHT}28                      & {\LIGHT}0                      & {\LIGHT}43                    &&
                             {\LIGHT}24                      & {\LIGHT}11                     & {\LIGHT}19                    &&
                             {\LIGHT}16\%                    & {\LIGHT}20\%                   & {\LIGHT}54\%                  \\
\multicolumn{1}{l}{{\IT}+any} &{\LIGHT}131                  & {\LIGHT}33                     & {\LIGHT}91                    &&
                             {\LIGHT}28                      & {\LIGHT}8                      & {\LIGHT}52                    &&
                             {\LIGHT}24                      & {\LIGHT}15                     & {\LIGHT}18                    &&
                             {\LIGHT}16\%                    & {\LIGHT}14\%                   & {\LIGHT}72\%                  \\  
& \multicolumn{3}{c} {(e) MRE IQR}    && \multicolumn{3}{c} {(f) Pred40 IQR}   &&\multicolumn{3}{c} {(g) SA IQR}      && \multicolumn{3}{c} {(h) D2H IQR} \\
~\\
\multicolumn{16}{c}{\small{{\bf Table \ref{tab:RQ3}.b: IQR results (over 20 runs).}}}\\~\\

        \end{tabular}
          \caption{RQ3 results.  Same format at Table~\ref{tab:RQ1}; i.e. 
     {\em Darker} cells show {\em better} results.  
    }
        \label{tab:RQ3}
\end{table}

One way to summarize   Table~\ref{select}  is that all these operators add a second round of landscape analysis
to the landscape analysis described in  \S\ref{aboutit}. 
Note that some of these SELECT operators require
more evaluations that others (``ALL'' uses the most while ``ANY'' uses the least).

The results of Table~\ref{tab:RQ3} let us comparatively assess these different SELECT operators.  
As might be expected the ``ANY'' method
performs worst since it evaluates the least. Interestingly, the ``SALL'' and ``ALL'' methods seem competitive.
This is a useful finding since, as shown in  
Table~\ref{tab:globmedian},
``SALL'' performs far fewer evaluations than ``ALL''.
Note that runtime difference in the two methods: 10 and 487 seconds
for ``SALL'' and ``ALL'' (respectively).  We conjecture that ``SALL'' does so well since the recursive tree methods we use for our landscape analysis (including the ``SALL'' method) 
are powerful enough to uncover   important features of our data. 

The last column of  Table~\ref{tab:globmedian} is a {\em Rank} measure 
computed via by   applying  the Ziztler's predicate (from \S\ref{better})
to the rows of  Table~\ref{tab:globmedian}. In a result consistent with our above notes, Ziztler prefers {IT} (with either ``SALL'' or ``ALL'') to everything else studied here. That said, and even though ``ALL'' reports a slightly better PRED40 and SA than ``SALL'', our engineering judgment is that we would prefer ``SALL'' since it is far faster.

\begin{table}
\caption{Runtime and performance comparison. (Medians on all datasets and all features)}
\small
\begin{center}
\begin{tabular}{l@{~}|c@{~}|c@{~}|c@{~}|c@{~}|c@{~}|c@{~}|c}
\textbf{Method}                 & \textbf{\#models}      & \textbf{MRE}               & \textbf{PRED40}            & \textbf{SA}       & \textbf{D2H}          & \textbf{Runtime (s)}                     & \textbf{Rank}   \\ \hline
\rowcolor[HTML]{EFEFEF}{\IT}+Sall           & 115       & 33         & 60         & 33       & 5.00\%       & 10     & 1      \\
{\IT}+all                                   & 6068      & 33         & 66         & 36       & 27.00\%      & 487    & 1      \\
RS                                          & 120       & 43         & 26         & 26       & 27.00\%      & 4      & 2      \\
{\SWAY}                                     & 44        & 52         & 26         & 26       & 45.00\%      & 94     & 2      \\
DE                                          & 132       & 132        & 6          & 6        & 50.00\%      & 4      & 3      \\
GS                                          & 2560      & 40         & 32         & 32       & 36.00\%      & 395    & 3      \\
FLASH                                       & 120       & 77         & 15         & 15       & 54.00\%      & 6      & 3      \\
{\IT}+Sany                                  & 66        & 118        & 6          & 6        & 63.00\%      & 7      & 4      \\
{\IT}+any                                   & 52        & 120        & 2          & 2        & 72.00\%      & 5      & 4      \\  
Default                                     & 1         & 70         & -5         & -5       & 50.00\%      & 1      & 4      \\
HYPEROPT                                    & 5000      & 35         & 7          & 0        & 77.00\%      & 237    & 4      \\
OPTUNA                                      & 3500      & 119        & 2          & 0        & 86.00\%      & 318    & 4  
\end{tabular}
\end{center}
\label{tab:globmedian}
\vspace{-14pt}
\end{table}

In summary, our answer to {\bf RQ3} is that :
\begin{RQ}
There is such a thing as too much landscape analysis.   ``SALL''  
is an appropriate compromise between   no landscape analysis and the excessive number of evaluations required for ``ALL''.  
\end{RQ}

\section{Threats to Validity}\label{threats}
As with any empirical study, biases can affect the final results. 
Therefore, any conclusions made from this work must be considered with the following issues in mind:

\textbf{Sampling Bias} - Any given empirical study using datasets suffers from sampling biases. 
In this work we have applied {\IT} to 12 different datasets containing OSS project health indicators, and to one learning algorithm (Random Forest). 
The behaviour of {\IT} upon different domains and different learners still needs to be evaluated and should be explored in future work.

\textbf{Parameter bias} - {\IT} contains many modular subprocesses within itself, containing different  hyperparameters. 
While this study shows that the current configuration of {\IT} is capable of producing results comparable to state-of-the-art algorithms, different configurations of the {\IT} system might prove valuable. 
That said, in defense of our current setup, we were able to find solutions that were significantly better than those found by state-of-the-art HPO techniques.

\textbf{Algorithm bias} - We have selected seven commonly applied and state-of-the-art HPO techniques to baseline {\IT} against. 
There might exist other HPO algorithms that were not explored in this study that may prove to be competitive to {\IT} in this domain, however the point of these baselines was to showcase the performance of {\IT} against other commonly applied techniques that are proven to obtain good results.

Finally, for this kind of optimization study, it is import to  address the following threat to validity. 
Any study that learns configurations from local data could be guilty of over-fitting, or worse, using the target data as part of the tuning work. To address that concern, we assert here that in all our studies, we (a)~sort the data by time stamps; then (b)~tune using past data to (c)~test on future data. 
That is, {\em our test data is never used as part of the tuning process}.

\section{Conclusion}
This paper has  explored open source software project health prediction.
As shown by our {\bf RQ1} results, when data is scare, learners may generate models with high error rates.
 Traditionally,
this problem has been mitigated by either
 taking the time to collect more data or artificially generate more data~\cite{chawla2002smote}.
 Here we show that, for the specific case of predicting project health for OSS projects, it is possible
 to address data scarcity by just some fine-tuning of pre-existing learners (in our case, random forests).

But this kind of {\em hyperparameter optimization}
can be like
walking around a broken car and randomly
hitting it till it works. Our approach, called {\IT}, adopts a landscape analytics approach that looks at the shape of
the data before deciding to move one way or another.
% We have applied {\IT} to the open-source software projects health domain, to aid in the prediction of indicators of project health one year into the future. 
% Previous work in this arena has obtained good results~\cite{xia2018hyperparameter} but suffered
% from high prediction errors. 
% %Nevertheless, we  have argued for the necessity of more accurate predictions in this realm, and as such, better optimized learners are a necessity so that decision makers are empowered with the tools for managing OSS projects, deciding on it's adoption and many other tasks.
% {\IT} can significantly reduce those errors such that they satisfy the Sarro limits (discussed 
% in our introduction).
As shown in our {\bf RQ2} results, 
when compared to    other state-of-the-art HPO techniques,
{\IT}  performs better  by a statistically
significantly amount.
Further, it achieves significantly better results after evaluating a fraction of the models needed by several hyperparameter optimizations.
When compared to a current state-of-the-art optimizer
(HYPEROPT and  OPTUNA):
\be
\item {\IT} is orders of magnitude faster
\item {\IT}  only evaluates around 2\% 
(115/5000) of the models needed by other methods.
\ee
This suggests that landscape analytics might be   applicable for more expensive evaluation functions in the future (i.e. deep learning).  We conjecture that {\IT}  does so well since the recursive tree methods we use for our landscape analysis   
are powerful enough to uncover   important features of our data. 

Finally, in our {\bf RQ3} results, we asked if there was such a thing as too much landscape analysis. We find that
our {\em Sall} policy is a good balance between too little and too much investigation of the shape of the data.

Based on the above,  we recommend   landscape analytics (e.g. {\IT}) especially  when learning from very small data sets.
While this paper has only explored the application of {\IT} to project health,
we see nothing in principle that
prevents the application of this technique to a wider range of problems.

% Given these results, we recommend  {\IT} is a useful approach for  hyperparameter optimization
% in the software engineering data domain, and should be considered by future work when there is a need for tuning learners
% (especially when the evaluation function of these learners is time consuming). 
% As said in our introduction, these results strongly recommend landscape analytics like {\IT} for ``data starved'' SE
% domains. To date, we have only applied landscape analytics to     predicting open source project health. But we see nothing in principle that prevents the
% application of this technique to a wider range of problems.
%We also suggest that it could be insightful to apply these methods to different domains, and pairing it with (say) deep learning algorithms.

\section*{Acknowledgments}
This work was partially supported by an 
NSF CCF award \#1908762 

\section*{Conflict of Interest Statement}
The authors declared that they have no conflict of interest.
\section*{Data Availability Statement}
All our data and scripts are on-line at
\url{https://github.com/zxcv123456qwe/niSneak}.
Permission is  granted, free of charge, to any person obtaining a copy
of this software and associated documentation files (the ``Software''), to deal
in the Software without restriction, including without limitation the rights
to use, copy, modify, merge, publish, distribute, sublicense, and/or sell
copies of the Software, and to permit persons to whom the Software is
furnished to do so, subject to the   
conditions of the MIT License.

{\footnotesize \bibliographystyle{ACM-Reference-Format}
\bibliography{bibliography.bib,other.bib,patrick.bib}
}

\newpage
\newcommand{\BLUE}[1]{\textcolor{blue}{#1}}

\end{document}